\documentclass[10pt]{article}
\usepackage{amsfonts}
\usepackage{amsmath}
\usepackage{amssymb}
\usepackage{epsfig}
\usepackage{graphicx,color}
\usepackage{mathptmx}
\def\l{\ell}
\def\sh{\hat{s}}
\def\mh{\hat{m}}

\def\mvh{\mh_{\phi}}

\def\mlh{\mh_\l}

\def\uh{\hat{u}}
\begin{document}
\title{Study of $B_s \to \phi \ell^+\ell^-$ Decay in a Single Universal Extra Dimension}
\author{Ying Li\footnote{Email: liying@ytu.edu.cn},~~~~ Juan Hua,\\
{\it  Department of Physics, Yantai University, Yantai 264-005,
China}}
 \maketitle
\begin{abstract}
Utilizing  form factors calculated within the light-cone sum rules,
we have evaluated the decay branching ratios of $B_s\to \phi\gamma$
and $B_s\to \phi \ell^+\ell^-$ in  a single universal extra
dimension model (UED), which is viewed as one of the alternative
theories beyond the standard model (SM). For the decay $B_s \to \phi
\ell^+\ell^-$, the dilepton invariant mass spectra, the
forward-backward asymmetry, and double lepton polarization are also
calculated. For each case, we compared the obtained results with
predictions of the SM. In lower values of the compactification
factor $1/R$, the only parameter in this model, we see the
considerable discrepancy between the UED and SM models. However,
when $1/R$ increases, the results of UED tend to diminish and at
$1/R = 1000~\mathrm{GeV}$, two models have approximately the same
predictions. Compared with data from CDF of $B_s \to \phi \mu^+
\mu^-$, the $1/R$  tends to be larger than $350~\mathrm{GeV}$. We
also note that the zero crossing point of the forward-backward
asymmetry is become smaller, which will be an important plat to prob
the contribution from the extra dimension model. The results
obtained in this work will be very useful in searching new physics
beyond SM. Moreover, the order of magnitude for branching ratios
shows a possibility to study these channels at the Large Hadron
Collider (LHC), CDF and the future super-B factory.
\end{abstract}

\section{Introduction}
As one kind of the flavor changing neutral current processes, the
electroweak penguin decays $b\to s\ell^+\ell^-$ appearing only at
the loop level in the standard model (SM), are therefore sensitive
to the fine structure of SM and to the  possible new physics as
well, and are expected to shed light on the existence of new physics
before the possible new particles are produced at colliders. With
the data from Belle and BaBar experiments, the exclusive processes
$B \to K^{(*)}\ell^+\ell^-$ have received great interest so that
their theoretical calculation has been the subject of many
investigations in the SM \cite{SM_ANA,Ali:1999mm} and beyond
\cite{NP_ANA,ued}. Along this line, the exclusive decays $B_s \to
\phi \ell^+\ell^-$ become also attractive since these decays are
also induced by $b\to s\ell^+\ell^-$, and could be measured at the
running Tavatron, LHC and future super-B factories. Recently, the
CDF collaboration had observed the rare semi-leptonic decay $B_s \to
\phi \mu^+\mu^-$ \cite{CDFPhill} and the branching ratio is
\begin{eqnarray}
\mathrm{Br}(B_s \to \phi \mu^+\mu^-)=\left[1.44\pm
0.33(\mathrm{stat.})\pm 0.46(\mathrm{syst.})\right]\times 10^{-6}.
\end{eqnarray}
This exclusive process is quite worthy of intensive research and
have attached much attention \cite{bsphill,bsphiued}. When studying
the semi-leptonic decays, dilepton invariant mass spectrum, the
forward-backward asymmetry, and double lepton polarization are
important observables to test SM and prob new physics, while the
first two are mostly analyzed. Due to small mass of electron and
muon, the invariant mass spectra and branching ratios are almost the
same for electron and muon modes. Meanwhile, it is very difficult to
measure the electron polarization, so we only consider $B_s \to \phi
\mu^+\mu^-, \tau^+\tau^-$ in this work.

To make the theoretical predictions clearly, additional knowledge of
decay form factors is needed, which is related with the calculation
of hadronic transition matrix elements, and can only be reliably
calculated by using a non-perturbative QCD method. Fortunately,
these form factors related to $B_s \to \phi \ell^+\ell^-$ have been
explored within the different methods, such as light cone sum rules
(LCSRs) \cite {Ball1998, Ball2005, Wu2006}, perturbative QCD
approach \cite{Wang:2007an} and in different quark models:
relativistic constitute quark model \cite{Melikov2000}, constituent
quark model \cite{Deandrea}, light front quark model
\cite{Geng2003}. Among them, the light-cone sum rules, which deal
with form factors at small momentum region, is complementary to the
lattice approach and has consistence with perturbative QCD and the
heavy quark limit. We will adopt the form factors calculated by the
LCSRs.

Among the ideas proposed to extend the SM, a lot of attention has
recently been devoted to models including extra dimensions
\cite{arkani,Randall:1999ee}. An interesting model is that proposed
by Appelquist, Cheng and Dobrescu with the so-called universal extra
dimensions (UED) \cite{ACD}, which means that all the SM fields may
propagate in one or more compact extra dimensions. The
compactification of the extra dimensions involves the appearance of
an infinite discrete set of four dimensional fields which create the
so-called KK particles. The simplest UED scenario is characterized
by a single extra dimension. Compared to the SM, this model has one
extra parameter called compactification radius, $R$. Hence, this
model is a minimal extension of the SM in $4 + 1$ dimensions with
the extra dimension compactified to the orbifold $S^1 /Z_2$ and the
fifth coordinate, $y$ running between $0$ and  $2\pi R$, and $y = 0$
and $y = \pi R$ are fixed points of the orbifold. The zero  modes of
fields propagating in the extra dimension correspond the SM
particles. The masses of KK particles are related to
compactification radius according to the relation $m_{n}^2 = m^2_0 +
n^2/ R^2$ , with $n = 1,2, ...$. One of the important property of
the model is the conservation of KK parity that guarantees the
absence of tree level KK contributions to low energy processes
occurring at scales much smaller than the compactification scale.

After the UED model being proposed, many attempts have been done to
constraint the only parameter compactification radius $R$, for
example, from Tevatron experiments the bound on the inverse
compactification radius is found to be about $1/R\geq 300
\mathrm{GeV}$. The anomalous magnetic moment of muon and $Z \to \bar
b b$ vertex  also lead to the same conclusion. Rare $B$ transitions
can also be used to constrain this scenario. Buras and collaborators
\cite{buras} have investigated the impact of universal extra
dimensions on the $B^0_{d,s}- \bar B^0_{d,s}$ mixing mass
differences, on the CKM unitarity triangle and on inclusive $b\to s$
decays for which they have computed the effective Hamiltonian. In
particular, it was found that $BR(B\to X_s\gamma$) allowed to
constrain $1/R > 250 \mathrm{GeV}$, a bound updated by a more recent
analysis to $1/R > 600 \mathrm{GeV}$ at $95\%$ CL, or to $1/R > 330
\mathrm{GeV}$ at $99\%$ CL \cite{Haisch:2007vb}. In this work, we
will consider the $1/R$ from $200 \mathrm{~GeV}$ up to $1000
\mathrm{~GeV}$. In the past years, the UED model has been applied
widely to calculate many observables related to the radiative and
semileptonic decays of hadrons (see for example
\cite{ued,bsphiued,azizi1,colangelo,aliev,aslambey}).

The aim of the paper is to find the effects of the KK modes on
various observables related to the $B_s \to \phi \ell^+\ell^-$
transition, and these observables involve the dilepton invariant
mass spectra, the forward-backward asymmetry, and double lepton
polarization. We will compare the obtained results with the
predictions of the standard model. In Ref.~\cite{bsphiued}, R.
Mohanta and A. K. Giri had calculated the processes $B_s \to \phi
\ell^+\ell^-$ and the $B_s\to \gamma \ell^+\ell^-$ under the UED
model, where they calculated the branching ratios and
forward-backward asymmetries adding the long distance contribution.
In this work, we will drop the contribution from the resonances,
such as $J/\psi, \psi^{\prime}$, and recalculate all observables.
Moreover, we will also calculate the lepton polarizations, which are
always viewed as good places to prob the new physics contribution.
In other words, this work can be regarded as supplementary of
Ref.\cite{bsphiued}.

The outline of the paper is as follows. In section 2, after
introducing the effective Hamiltonian responsible for the $b\to s
\ell^+\ell^-$ transition and form factors of $B_s \to \phi$, we will
present the formula of observable. In section 3,  we numerically
analyze the considered observables of $B_s  \to \phi \mu^+ \mu^-$,
$B_s \to \phi \tau^+ \tau^-$. This section also includes a
comparison of the results obtained in UED model with that predicted
by the SM. We will summarize this work at last.
\section{Effective Hamiltonian, Form Factors and Formula of Observable}
At quark level, the $B_s  \to \phi \ell^+ \ell^-$ transition proceed
via FCNC transition of the $b \to s \ell^+ \ell^- $. Neglecting the
doubly Cabibbo-suppressed contributions, the effective Hamiltonian
governing $b\to s \ell^+\ell^-$ transition is given
by~\cite{Altmannshofer:2008dz,Chetyrkin:1996vx}
\begin{equation} \label{eq:Heff}
{\cal H}_{ eff} = - \frac{4 G_F}{\sqrt{2}}V_{tb}V_{ts}^{*}
\sum_{i=1}^{10} C_i(\mu) O_i(\mu) \,,
\end{equation}
where explicit expressions of $O_{i}$ could be found in
Ref.~\cite{Altmannshofer:2008dz}, and the Wilson coefficients $C_i$
can be calculated perturbatively
\cite{Beneke:2001at,bobeth,bobeth02,Huber:2005ig}. Using the
effective Hamiltonian, the free quark decay amplitude can be written
as:
\begin{eqnarray}
{\cal M} &=& {G_F \alpha_{em} V_{tb} V_{ts}^\ast \over 2\sqrt{2}
\pi} \Bigg[ C_9^{eff} \bar{s}\gamma_\mu (1-\gamma_5) b \, \bar{\ell}
\gamma^\mu \ell + C_{10}  \bar{s} \gamma_\mu (1-\gamma_5) b \,
\bar{\ell} \gamma^\mu
\gamma_5 \ell \nonumber \\
& &  -2 m_b C_7^{eff}  {{q^\nu}\over q^2} \bar{s} i \sigma_{\mu\nu}
(1+\gamma_5) b \, \bar{\ell} \gamma^\mu \ell \Bigg]~.
\end{eqnarray}
Here $q=p_++p_-$, where $p_\pm$ are the four momenta of the leptons,
respectively. Noted that ${\cal M}$, although a free quark decay
amplitude, contains part of long-distance effects from four-quark
interactions, which usually are absorbed into a redefinition of the
short distance Wilson coefficients. To be specific, we define the
effective coefficient of the operator $O_9$ as£º
\begin{eqnarray}
C_9^{eff}=C_9+Y(q^2),
\end{eqnarray}
where $Y(q^2)$ stands for the above mentioned contribution from
four-quark interaction. The corresponding  operators and $Y(q^2)$
can refer to Ref.~\cite{Buras:1993xp}. In this work, as mentioned
before, we have not consider the contribution mainly due to $J/\psi$
and $\psi^\prime$ resonances in the decay chain $B_s\to \phi
\psi^{(\prime)}\to \phi \ell^+\ell^-$, which could be vetoed
experimentally.

\begin{figure}[thb]
\begin{center}
\includegraphics[scale=0.7]{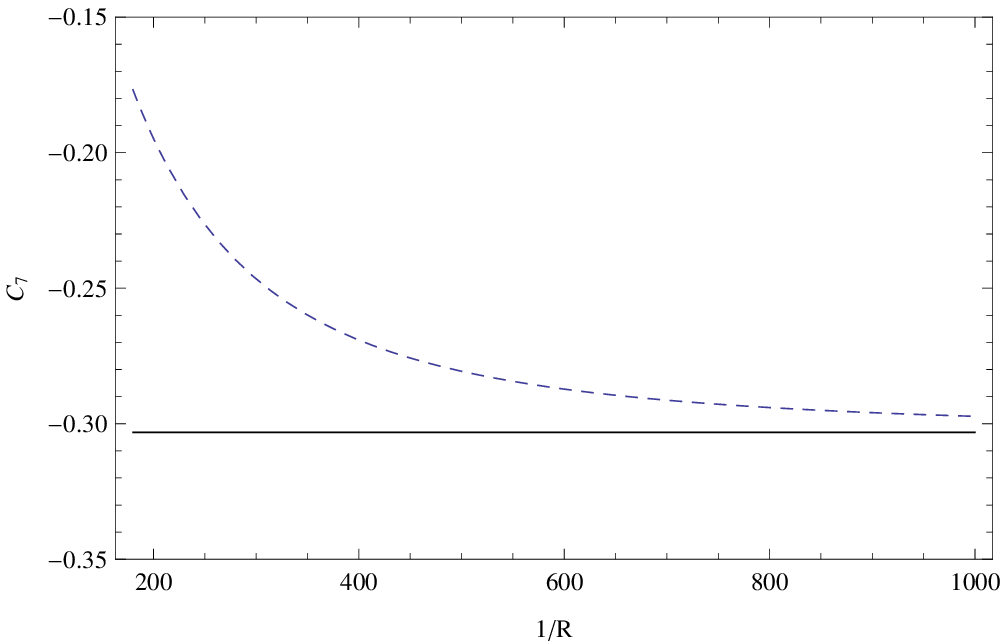}\,\,\,
\includegraphics[scale=0.7]{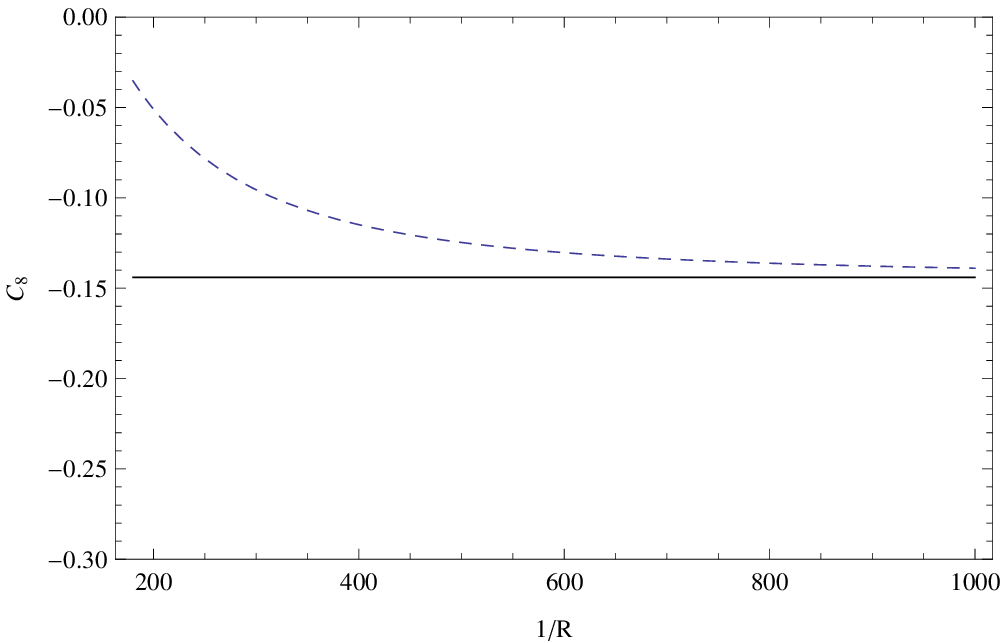}\\
\includegraphics[scale=0.7]{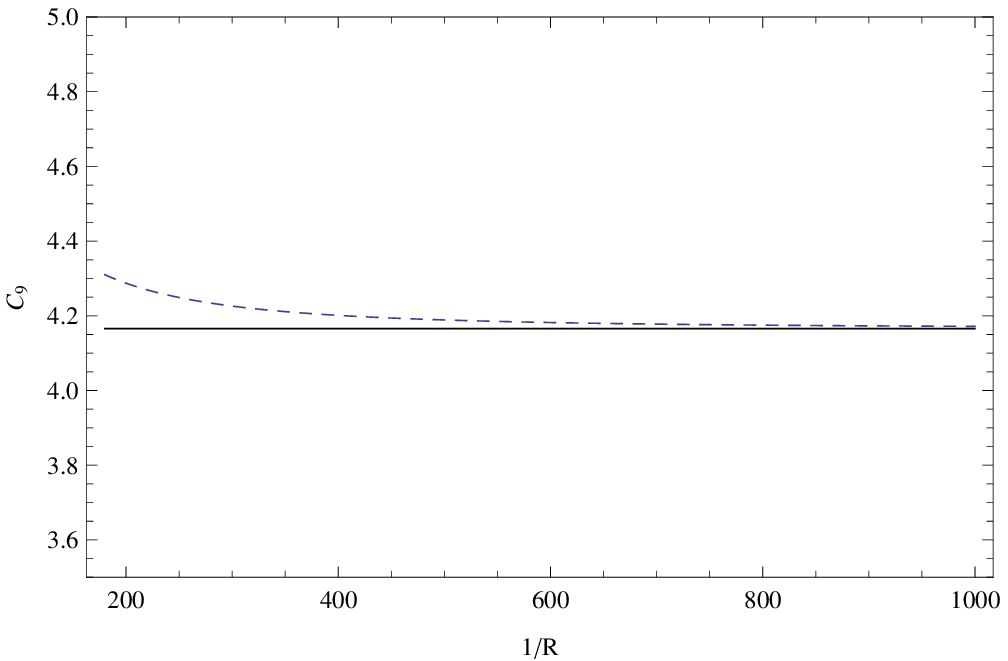}\,\,\,
\includegraphics[scale=0.7]{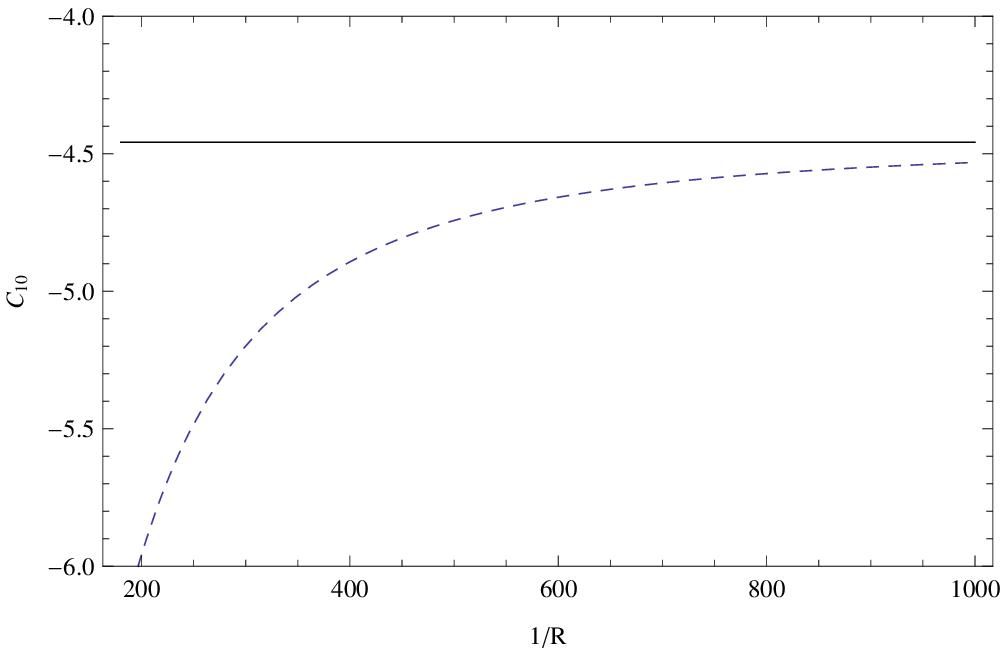}
\caption{The wilson coefficients ($\mu=4.8 \mbox{GeV}$) plotted
versus $1/R$. The constant lines are the SM values. } \label{Fig:1}
\end{center}
\end{figure}

The main source of the deviation of the UED model and  SM
predictions on the considered observables are  from Wilson
coefficients $C_7^{eff}$, $C_9^{eff}$ and $C_{10}$, which can be
expressed in terms of the periodic functions, $F(x_{t},1/R)$ with
$x_{t}=m_{t}^{2}/M_{W}^{2}$ and $m_t$ being the top quark mass.
Similar to the mass of the KK particles described in terms of the
zero modes $(n = 0)$ correspond to the ordinary particles of the SM
and additional parts coming from the UED model, the functions,
$F(x_{t},1/R)$ are also  written in terms of the corresponding SM
functions, $F_0 (x_t )$ and  additional parts which are functions of
the compactification factor, $1/R$, i.e.,
\begin{eqnarray}
F(x_t,1/R)=F_0(x_t)+\sum_{n=1}^{\infty}F_n(x_t,x_n),
\end{eqnarray}
where $x_n=\displaystyle{m_n^2 \over M_W^2}$ and
$m_n=\displaystyle{n \over R}$. The Glashow-Illiopoulos-Maiani (GIM)
mechanism guarantees the finiteness of the functions, $F(x_t,1/R)$
and satisfies the condition, $F(x_t,1/R)\to F_0(x_t)$, when $R\to
0$. As far as $1/R$ is taken in the order of a few hundreds of
$GeV$, the Wilson coefficients differ considerably from the SM
values. For explicit expressions of the Wilson coefficients in UED
model see \cite{ued,buras}. In Fig.~\ref{Fig:1}, we plot the Wilson
coefficients $C_i \,(i=7,8,9,10)$ versus $1/R$ and find that the
impact of the UED on the $C_9$ is small. The suppression of $|C_7|$
and $|C_8|$, that for $1/R=300~ \mathrm{GeV}$ amount to $82\%$ and
$66\%$ relative to the SM values, respectively. For $|C_{10}|$, it
can be enhanced by $16\%$ for $1/R=300~ \mathrm{GeV}$, which does
not renormalize under QCD. The UED contribution to four quark QCD
penguin operators  are also neglected in this work because of rather
smaller Wilson coefficients.

For the process $B_s\to  \phi \ell^+\ell^-$, the  nonvanishing
matrix elements are:
\begin{eqnarray}
\langle  \phi(k) | \bar s\gamma_\mu(1-\gamma_5) b | \bar
B_s(p)\rangle  &=& -i \epsilon^*_\mu (m_{B_s}+m_{\phi}) A_1(q^2) + i
(2p-q)_\mu (\epsilon^* \cdot q)\,
\frac{A_2(q^2)}{m_{B_s}+m_{\phi}}\, \nonumber \\
& &   +i q_\mu (\epsilon^* \cdot q) \, \frac{2m_{\phi}}{q^2}\,
\Big[A_3(q^2)-A_0(q^2)\Big] \, +
\epsilon_{\mu\nu\rho\sigma}\epsilon^{*\nu} p^\rho k^\sigma\,
\frac{2V(q^2)}{m_{B_s}+m_{\phi}}\,,
\end{eqnarray}
\begin{eqnarray}
\langle \phi(k) | \bar s \sigma_{\mu\nu} q^\nu (1+\gamma_5) b |
\bar{B}_s(p)\rangle &=& i\epsilon_{\mu\nu\rho\sigma} \epsilon^{*\nu}
p^\rho k^\sigma \, 2 T_1(q^2)\,  + T_2(q^2) \Big[\epsilon^*_\mu
(m_{B_s}^2-m_{\phi}^2)
- (\epsilon^* \cdot q) \,(2p-q)_\mu \Big] \, \nonumber\\
&+ &  T_3(q^2) (\epsilon^* \cdot q) \left[q_\mu -
\frac{q^2}{m_{B_s}^2-m_{\phi}^2}\, (2p-q)_\mu \right]\,,
\end{eqnarray}
where $\epsilon_\mu$ is the polarization vector of the $\phi$ meson.
By means of equation of motion, one can obtain several relations
between form factors
\begin{eqnarray}
A_3(q^2) = \frac{m_{B_s}+m_{\phi}}{2m_{\phi}}\, A_1(q^2) -
\frac{m_{B_s}-m_{\phi}}{2m_{\phi}}\, A_2(q^2)
\end{eqnarray}
and $A_0(0) =  A_3(0)$,$T_1(0) =  T_2(0)$. All sighs are defined in
such a way to render the form factors are real and positive. We will
use the results calculated  by using the technique of the light-cone
QCD sum rule approach~\cite{Ball1998}, the updated results and $q^2$
dependence of the form factors can be found in Ref.~\cite{Ball2005}.
In Ref.\cite{bsphiued}, the authors found that the uncertainties
from the form factors can change slightly from their corresponding
central values in the low $s$ region, and highly suppressed in large
$s$ region, we will not consider the effect of these uncertainties
here. The physical range in $s=q^2$ extends from $s_{\rm min} = 4
m_l^2$ to $s_{\rm max} =(m_{B_s}-m_{\phi})^2$.

Keeping the lepton mass and adopting the same convention and
notation as \cite{Ali:1999mm},  we find that the dilepton invariant
mass spectrum for $\bar{B}_s\to \phi\ell^+\ell^-$ decay is given as
\begin{eqnarray}
\frac{\mathrm{d}\Gamma}{\mathrm{d}\sh} = \frac{G_F^2 \,
\alpha_{em}^2 \, m_{B_s}^5}{2^{10} \pi^5}
      \left| V_{ts}^\ast  V_{tb} \right|^2 \, \uh(\sh) \,  D \;,
\end{eqnarray}
where the function $D$ is defined as:
\begin{multline}
D  =  \frac{|A|^2}{3} \sh \lambda (1+2 \frac{\mlh^2}{\sh}) +|E|^2
\sh \frac{\uh(\sh)^2}{3}+\frac{1}{4 \mvh^2} \left[ |B|^2
(\lambda-\frac{\uh(\sh)^2}{3} + 8 \mvh^2 (\sh+ 2 \mlh^2))
          + |F|^2 (\lambda -\frac{ \uh(\sh)^2}{3} + 8 \mvh^2 (\sh- 4 \mlh^2))
\right]\Bigg. \\
    +\Bigg.\frac{\lambda }{4 \mvh^2} \left[ |C|^2 (\lambda - \frac{\uh(\sh)^2}{3})
 + |G|^2 \left(\lambda -\frac{\uh(\sh)^2}{3}+4 \mlh^2(2+2 \mvh^2-\sh) \right)
\right]
   -\frac{1}{2 \mvh^2}
\left[ {\rm Re}(BC^\ast) (\lambda -\frac{ \uh(\sh)^2}{3})(1 - \mvh^2
- \sh)
 \right. \Bigg.\\
  + \left.  \Bigg.{\rm Re}(FG^\ast) ((\lambda -\frac{ \uh(\sh)^2}{3})(1 - \mvh^2 - \sh) +
4 \mlh^2 \lambda) \right]
  -2 \frac{\mlh^2}{\mvh^2} \lambda  \left[ {\rm Re}(FH^\ast)-
 {\rm Re}(GH^\ast) (1-\mvh^2) \right] +\frac{\mlh^2}{\mvh^2} \sh \lambda |H|^2    \,.
 \label{eq:ims}
\end{multline}
With $\sh=s/m_B^2$, $\mlh=m_l/m_B$ and  $\mvh=m_\phi/m_B$, the
kinematic variables are defined as
\begin{eqnarray}
\lambda &=&1+\mvh^4+\sh^2-2 \sh- 2\mvh^2(1+\sh)         \\
\uh(\sh)&=&\sqrt{\lambda (1- 4\frac{\mlh^2}{\sh})}
\end{eqnarray}
Combined the effective coefficients and  form factors, the auxiliary
functions $A\,,B\,,C\,,E\,,F\,,G$ and $H$ are refereed to Ref.
\cite{Ali:1999mm}. According to the definition of the
forward-backward asymmetry (FBA), it is straightforward to obtain
the expression of the normalized FBA as:
\begin{eqnarray}\label{EqAFB}
  \frac{\mathrm{d} A_{\rm FB}^{\phi}}{\mathrm{d} \sh}D=\uh(\sh)
   \sh \left[{\rm Re}(BE^\ast)+{\rm Re}(AF^\ast)
  \right]
\end{eqnarray}

We define the three orthogonal unit vectors in the center mass frame
of dilepton as
\begin{eqnarray}
      \hat{e}_L=\vec{p}_+,\,\,\,\,\,\,
      \hat{e}_N=\frac{\vec{p}_K \times \vec{p}_+ }{|\vec{p}_K \times
      \vec{p}_+|},\,\,\,\,\,\,
      \hat{e}_T= \hat{e}_N \times \hat{e}_L\; ,
\end{eqnarray}
which are related to the spin of lepton by a Lorentz boost.
  Then, the decay width of the $B_s \to\phi \ell^+ \ell^-$ decay for any spin direction
$\hat{n}$ of the lepton, where $\hat{n}$ is a unit vector in the
dilepton center mass frame, can be written as:
\begin{eqnarray}
      \frac{d\Gamma(\hat{n})}{d\hat{s}}=\frac{1}{2}\big (\frac{d\Gamma}{d\hat{s}}\big )_0[1
      +(P_L\hat{e}_L+P_N\hat{e}_N+P_T\hat{e}_T)\cdot\hat{n}]
\end{eqnarray}
where the subscript $"0"$ denotes the unpolarized decay width, $P_L$
and $P_T$  are the longitudinal and transverse polarization
asymmetries in the decay plane respectively, and $P_N$ is the normal
polarization asymmetry in the direction perpendicular to the decay
plane.

  The lepton polarization asymmetry $P_i$ can be obtained by calculating
\begin{eqnarray}
P_i(\hat{s})=\frac{d\Gamma(\hat{n}=\hat{e}_i)/d\hat{s}-
d\Gamma(\hat{n}=-\hat{e}_i)/d\hat{s}}{d\Gamma(\hat{n}=\hat{e}_i)/d\hat{s}+
d\Gamma(\hat{n}=-\hat{e}_i)/d\hat{s}}\; .
\end{eqnarray}
By a straightforward calculation, we get
 \begin{eqnarray}
 P_LD &=&\sqrt{1-4\frac{\hat{m}^2_l}{\hat{s}}}\left\{\frac{2\hat{s}\lambda}{3}
 \mathrm{Re}(AE^{\dagger})+\frac{(\lambda+12\hat{m}^2_{\phi}\hat{s})}{3\hat{m}^2_{\phi}}\mathrm{Re}(BF^{\dagger})\right.\nonumber\\
 &&\left.-\frac{\lambda(1-\hat{m}^2_{\phi}-\hat{s})}{3\hat{m}^2_{\phi}}\mathrm{Re}(BG^{\dagger}+CF^{\dagger})
 +\frac{\lambda^2}{3\hat{m}_{\phi}}\mathrm{Re}(CG^{\dagger})\right\} ,\\
 P_ND&=&\frac{-\pi\sqrt{\hat{s}}\hat{u}(\hat{s})}{4\hat{m}_{\phi}}\left\{
 \frac{\hat{m}_l}{\hat{m}_{\phi}}\left[\mathrm{Im}(FG^{\dagger})(1+3\hat{m}^2_{\phi}-\hat{s})\right.\right. \nonumber\\
 &&\Bigg.\left.+\mathrm{Im}(FH^{\dagger})(1-\hat{m}^2_{\phi}-\hat{s})-\mathrm{Im}(GH^{\dagger})\lambda  \right]+2\hat{m}_{\phi}\hat{m}_l
 [\mathrm{Im}(BE^{\dagger})+\mathrm{Im}(AF^{\dagger})]\Bigg\},\\
P_TD&=&\frac{\pi\sqrt{\lambda}\hat{m}_l}{4\sqrt{\hat{s}}}\Bigg\{
4\hat{s}\mathrm{Re}(AB^{\dagger})+\frac{(1-\hat{m}^2_{\phi}-\hat{s})}{\hat{m}^2_{\phi}}\left[
-\mathrm{Re}(BF^{\dagger})+(1-\hat{m}^2_{\phi})\mathrm{Re}(BG^{\dagger})+\hat{s}\mathrm{Re}(BH^{\dagger})\right]\Bigg.\nonumber\\
&&\left.+\frac{\lambda}{\hat{m}^2_{\phi}}[\mathrm{Re}(CF^{\dagger})-(1-\hat{m}^2_{\phi})\mathrm{Re}(CG^{\dagger})
-\hat{s}\mathrm{Re}(CH^{\dagger})]\right\}
\end{eqnarray}
\section{Numerical Analysis}
In this section, we will examine the above mentioned physics
observables and study their sensitivity to  the compactification
factor, $1/R$, which is the most important  parameter in the single
universal extra dimension model. Now, the parameters of the SM in
our calculation are listed as follows:
\begin{gather}
m_{B_s} = 5.36 \mbox{ GeV},\  m_{b} = 4.8 \mbox{ GeV},\ m_c = 1.4
\mbox{ GeV}, \  m_{\phi} = 1.02 \mbox{ GeV}, \  m_{\mu} = 0.1057
\mbox{ GeV}, \nonumber \\\  m_{\tau} = 1.7769 \mbox{ GeV} \
m_{t}=172.4\mbox{ GeV},\ m_{W}=80.4\mbox{ GeV},\ m_{Z}=91.18\mbox{
GeV},\ \sin^2 \theta_W
=0.23,\nonumber\\
\alpha_{em}=\frac{1}{137},\ \alpha_s(m_Z)=0.118, \
|V_{ts}^*V_{tb}|=38.5\times 10^{-3}, \
\tau_{B_s}=1.46\times10^{-12}\mbox{s}. \label{parameterZ}
\end{gather}
\begin{figure}[thb]
\begin{center}
\includegraphics[scale=0.7]{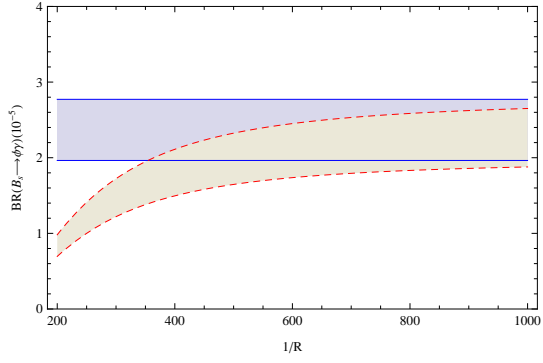}
\caption{The branching ratio of $B_s \to \phi \gamma$ changes with
$1/R$, and horizonal band corresponds to the SM prediction.}
\label{fig2}
\end{center}
\end{figure}

It is interesting to calculate the branching ratio of $B_s \to \phi
\gamma$ in the UED model firstly, since this channel will be
measured easily among the $B_s$ rare decays. Within the effective
Hamiltonian, see equation (\ref{eq:Heff}), the decay rate of this
channel is read:
\begin{eqnarray}
\Gamma(B_s \to \phi \gamma)=\frac{\alpha G_F^2 }{32
\pi^4}|V_{tb}V_{ts}^*|m_b^2m_{B_s}^3\left
(1-\frac{m_\phi^2}{m_{B_s}^2}\right )^3|C_7^{eff}|^2 |T_1(0)|^2
\end{eqnarray}
In the SM, we found that the $\mathrm{BR}(B_s \to \phi
\gamma)=(2.4\pm 0.4)\times 10^{-5}$, the major uncertainty is from
the  error of the form factor $T_1(0)$. Considering the contribution
of the UED, we present the  branching ratio dependence on
compactification parameter $1/R$ in Fig.~\ref{fig2}. There are
considerable discrepancies between the predictions of the UED and SM
models for low values of the $1/R$.  For $1/R=300~\mbox{GeV}$, the
ratio can reach to $ 1.5^{+0.2}_{-0.3} \times 10^{-5}$. Such a
discrepancy at low values of $1/R$ can be  a signal for the
existence of extra dimensions. Accordingly, if the experiment can
measure this channel well, we can constraint the range of $1/R$.

\begin{figure}
\begin{center}
\includegraphics[scale=0.7]{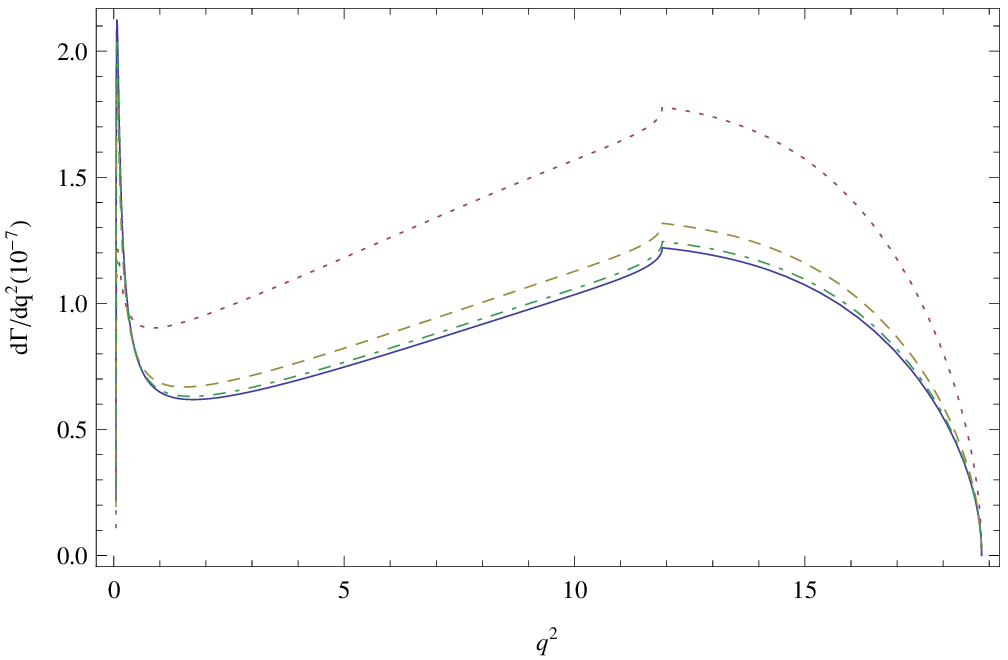}\,\,\,
\includegraphics[scale=0.7]{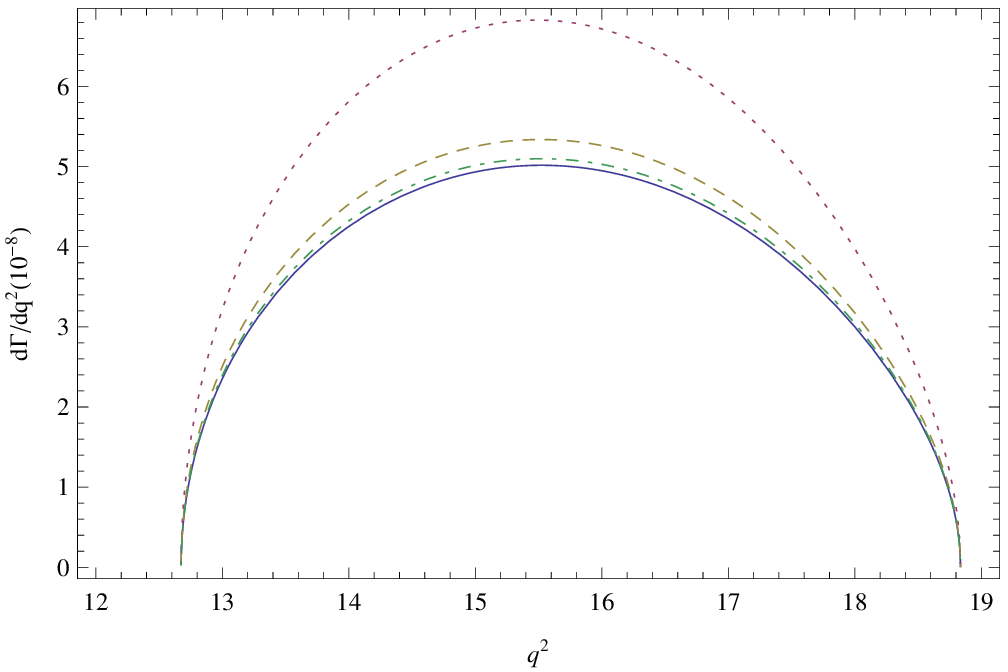}
\caption{The variation of the dilepton invariant mass spectra  of
$B_s \to\phi \mu^+\mu^-$(left panel) and $B_s \to\phi
\tau^+\tau^-$(right panel) with $q^2$ (in units of
$\mathrm{GeV}^2$). The solid line corresponds to the SM, dotted
line, dashed line and dot-dashed line are for $1/R=200~
\mathrm{GeV},500~ \mathrm{GeV}, 1000~ \mathrm{GeV}$ respectively.}
\label{Fig:3}
\end{center}
\end{figure}
\begin{figure}
\begin{center}
\includegraphics[scale=0.7]{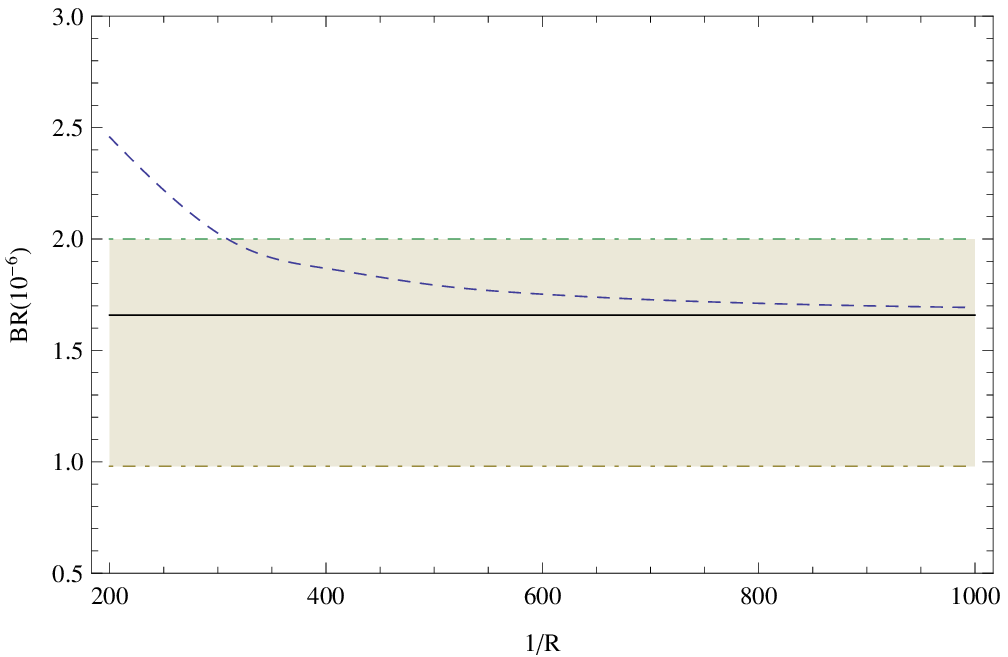}\,\,\,
\includegraphics[scale=0.7]{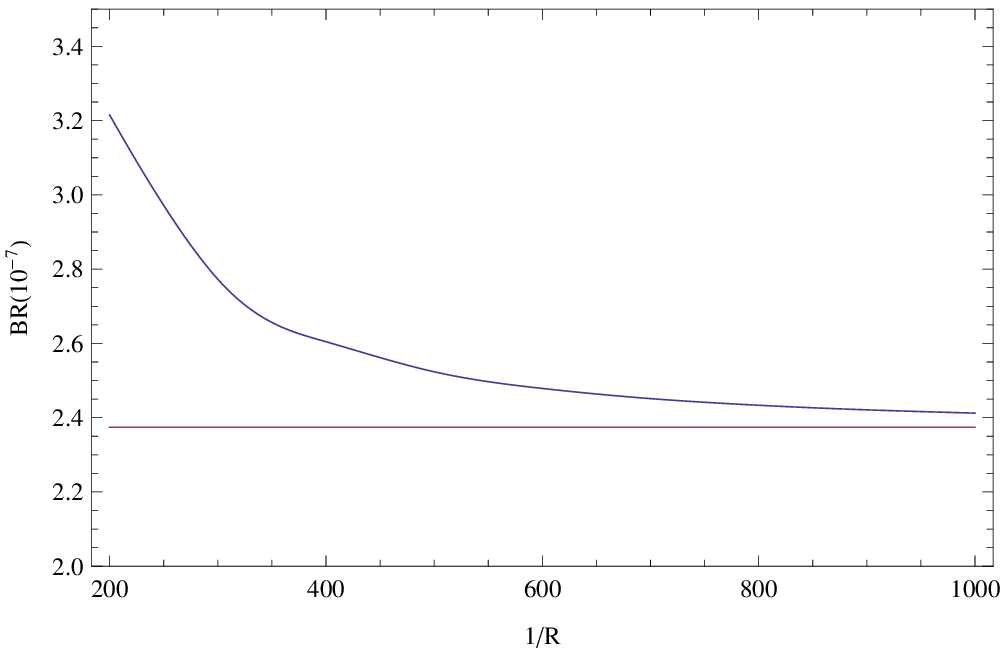}
\caption{The branching ratio of $B_s \to\phi \mu^+\mu^-$(left panel)
and $B_s \to\phi \tau^+\tau^-$(right panel)  changes with $1/R$, and
horizonal solid lines corresponds to the SM prediction. In the left
panel, the band area depicts the experimental data from CDF.}
\label{Fig:4}
\end{center}
\end{figure}
\begin{figure}
\begin{center}
\includegraphics[scale=0.9]{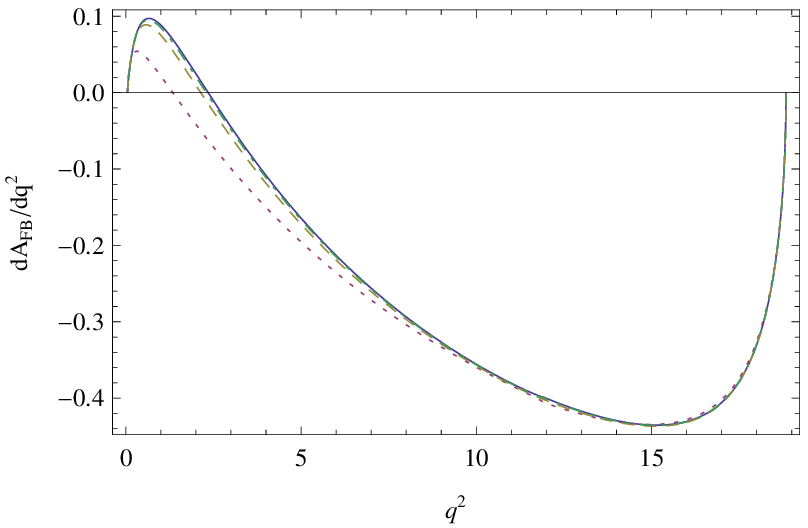}\,\,\,
\includegraphics[scale=0.9]{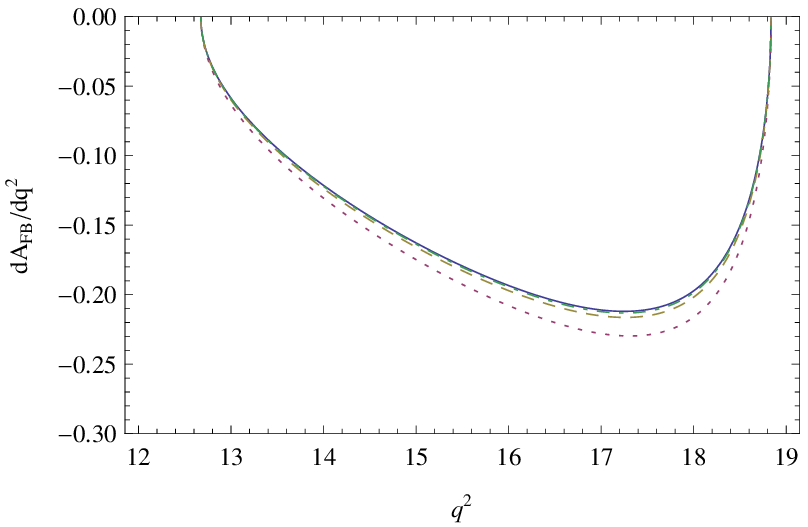}
\caption{The lepton forward-backward asymmetry  of $B_s \to\phi
\mu^+\mu^-$(left panel) and $B_s \to\phi \tau^+\tau^-$(left panel)
with $q^2$ (in units of GeV). The solid line corresponds to the SM,
dotted line, dashed line and dot-dashed line are for $1/R=200~
\mathrm{GeV},500~ \mathrm{GeV}, 1000~ \mathrm{GeV}$ respectively.}
\label{Fig:5}
\end{center}
\end{figure}
\begin{figure}
\begin{center}
\includegraphics[scale=0.7]{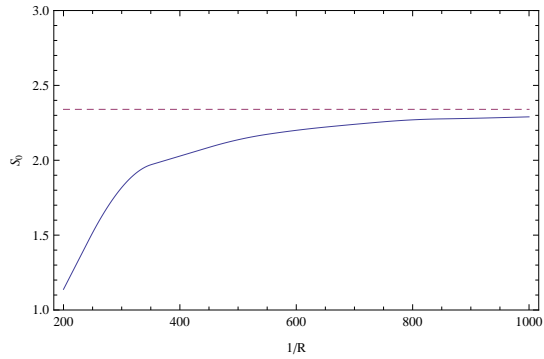}\,\,\,
\caption{The zero of the FBA changes with $1/R$, and horizonal line
corresponds to the SM prediction. } \label{Fig:6}
\end{center}
\end{figure}

In term of Eq.~(\ref{eq:ims}), we illustrate the dilepton invariant
mass spectra in Fig.~\ref{Fig:3}. By integrating the differential
ratios over $q^2$, in the SM, we obtain that:
\begin{eqnarray}
&&\mathrm{Br}(B_s \to \phi \mu^+\mu^- )= 1.8\times 10^{-6};\nonumber\\
&&\mathrm{Br}(B_s \to \phi \tau^+\tau^-)~~= 2.4\times 10^{-7}.
\end{eqnarray}
Considering the uncertainties in both theoretical side and
experimental sides, the prediction of $B_s \to \phi \mu^+\mu^-$ is
consistent with CDF measurement $\mathrm{Br}(B_s \to \phi \mu^+\mu^-
)= (1.44 \pm 0.56)\times 10^{-6}$  well, which is shown in
Fig.~\ref{Fig:4}. Note that the small branching ratio of tau mode is
due to highly suppressed phase space. We find there is  difference
between our results and these from the Ref.~\cite{bsphiued}, that is
because we ignore the resonance contribution and slight different
parameter space. Adding the the UED contribution, from the figures,
one can see that the extra dimension effects can increase both the
differential width and branching ratios for low values of $1/R$. As
$1/R$ increases, this difference tends to diminish so that for
higher values of $1/R$ ($1/R > 1000 \mathrm{GeV} $), the predictions
of UED become very close to the results of SM . Such discrepancy at
low values of $1/R$ can be considered as a signal for the existence
of extra dimensions. Moreover, once $1/R>500 \mathrm{GeV}$, it is
impossible to disentangle the extra dimension contribution clearly
in both modes. As it is expected, the order of magnitudes of the
branching ratios show a possibility to study such channels at the
LHC.

We show the lepton forward-backward asymmetries (FBA) for both
channels in Fig.~\ref{Fig:5}. It is shown that there is also
considerable discrepancy between the predictions of the UED and SM
models for low values of $1/R$.  As $1/R$ increases, this difference
starts to diminish. It should be stressed that the hadronic
uncertainties almost have no influence on this symmetry, so it can
provide the best tool to prob the new physics contribution. For the
muon mode, due to the destructive interference between the photo
penguin and $Z$ penguin, the FBA should have a zero point $s_0$. Any
deviation of the zero position $s_0$ from that of the SM  would give
us the clue of the new physics. In Fig.~\ref{Fig:5}, one can find
that $s_0$ decreases from the SM value. To make the situation
clearer, the dependence of $s_0$ on the compactification parameter
$1/R$ is depicted in Fig.~\ref{Fig:6}, which shows that the zero
position is pushed to a smaller value by decreasing $1/R$. Such a
sensitivity indicates $s_0$ is particularly suited to constrain the
$R$. Hence, with the enhancement of experimental precision and
statistics in LHC and CDF, the measurements of FBA would provide
more data and effectively pine the NP effects.

\begin{figure}
\begin{center}
\includegraphics[scale=1.0]{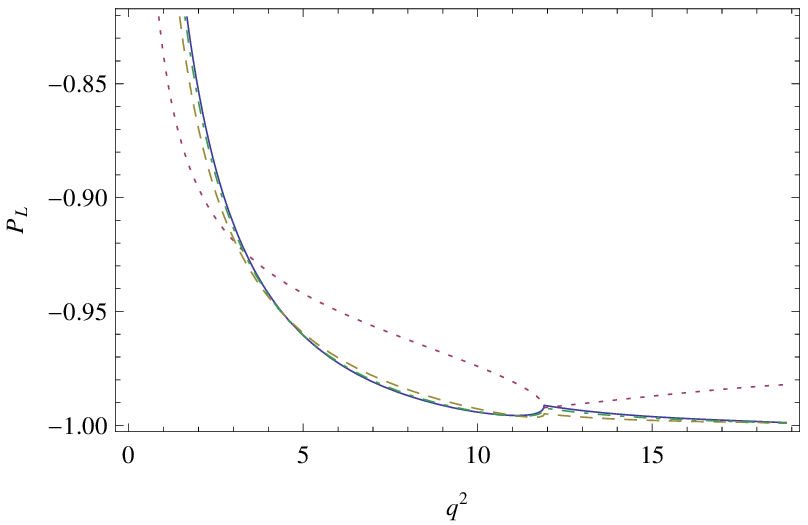}\,\,\,
\includegraphics[scale=1.0]{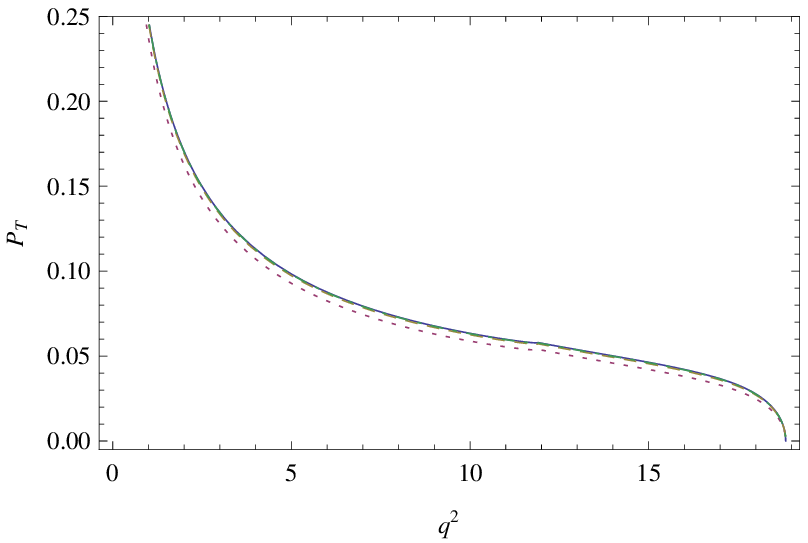}
\caption{The dependence  of longitudinal (left panel) and
transverse(right panel) polarization of $B_s \to\phi \mu^+\mu^-$
 with $q^2$ (in units of
$\mathrm{GeV}^2$). The solid line corresponds to the SM, dotted
line, dashed line and dot-dashed line are for $1/R=200~
\mathrm{GeV},500~ \mathrm{GeV}, 1000~ \mathrm{GeV}$ respectively. }
\label{Fig:7}
\end{center}
\end{figure}

\begin{figure}
\begin{center}
\includegraphics[scale=1.0]{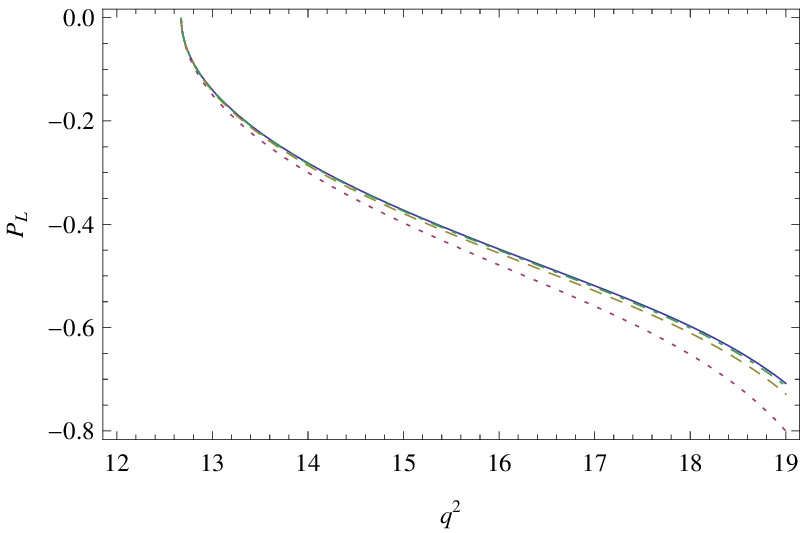}\,\,\,
\includegraphics[scale=1.0]{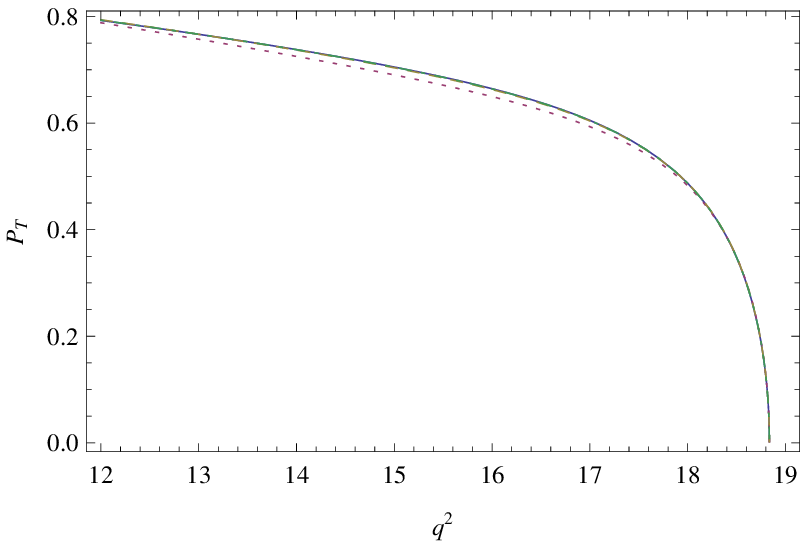}
\caption{The dependence  of longitudinal (left panel) and
transverse(right panel) polarization of $B_s \to\tau \mu^+\mu^-$
 with $q^2$ (in units of
$\mathrm{GeV}^2$). The solid line corresponds to the SM, dotted
line, dashed line and dot-dashed line are for $1/R=200~
\mathrm{GeV},500~ \mathrm{GeV}, 1000~ \mathrm{GeV}$ respectively.}
\label{Fig:8}
\end{center}
\end{figure}

Now, we turn to discuss the lepton polarization. In Fig.~\ref{Fig:7}
and ~\ref{Fig:8}, we present the longitudinal and transverse
polarization for $B_s \to\phi \mu^+\mu^-$ and $B_s \to\phi
\tau^+\tau^-$, respectively. For the normal polarization part, due
to  real $C_{10}$, it is the order of $10^{-3}$ and cannot be
observed even in the designed super-B factory, then we will ignore
this part in this work. From the Fig.~\ref{Fig:7} of $B_s \to\phi
\mu^+\mu^-$, for  small $1/R$, the deviation between UED model and
SM is apparent. At the small value of momentum transfer, the $|P_L|$
is larger than the SM prediction; while at $q^2>3.2 \mathrm{GeV}^2$,
it will become smaller than that of SM. Thus, measurement of $P_L$
will discriminate between the different models. For the $P_T$ part,
the predictions from different models are almost the same. As $B_s
\to\phi \tau^+\tau^-$ as concerned, with smaller
$1/R=200\mathrm{GeV}$, we can observe the deviation appears when
$q^2>16\mathrm{GeV}^2$. Again in the $P_T$ part the predictions are
the same as SM for all $1/R$.

\section{Summary}
Using form factors calculated within the light-cone sum rules, we
presented the decay branching ratios of $B_s\to \phi\gamma$ and
$B_s\to \phi \ell^+\ell^-$ in  a single universal extra dimension
model (UED), which is view as one of the alternative theories beyond
the standard model (SM) with only one new parameter. For the decay
$B_s \to \phi \ell^+\ell^-$, the dilepton invariant mass spectra,
the forward-backward asymmetries, and double lepton polarizations
are also calculated. For each case, we compared the obtained results
with predictions of the SM. In lower values of the compactification
factor $1/R$,  we can see the considerable discrepancy between the
UED and SM models. However, when $1/R$ increases, the results of UED
tend to diminish and at $1/R = 1000~\mathrm{GeV}$, two models have
approximately the same predictions. For $B_s \to \phi \mu^+\mu^-$,
compared with data from CDF, the $1/R$  tends to be larger than
$350~\mathrm{GeV}$. We also note that the zero point of the
forward-backward asymmetry  become smaller, which will be an
important plat to prob the contribution from the extra dimension.
Moreover, the order of magnitude for branching ratios shows a
possibility to study these channels at the Large Hadron Collider
(LHC), CDF and the future super-B factory.

\section*{Acknowledgement}
This work is supported in part by the NSFC (Nos.10805037 and
10625525) and the Natural Science Foundation of Shandong Province
(ZR2010AM036).  The authors would like to thank K. Azizi,  M. Jamil
and Qin Chang for helpful discussions and Cai-Dian L\"u for
improving the manuscript.


\begin{thebibliography}{99}
\bibitem{SM_ANA}
  For instance:
  N.~G.~Deshpande, J.~Trampetic, Phys.\ Rev.\ Lett.\ {\bf 60} (1988) 2583 ;
  W.~Jaus and D.~Wyler, Phys.\ Rev.\  D {\bf 41} (1990) 3405;
  G.~Burdman, Phys.\ Rev.\  D {\bf 52} (1995)  6400 [hep-ph/9505352];
  P.~Colangelo, F.~De Fazio, P.~Santorelli and E.~Scrimieri, Phys.\ Rev.\  D {\bf 53} (1996) 3672  [Erratum-ibid.\  D {\bf 57}  (1998) 3186] [hep-ph/9510403];
  C.~Q.~Geng and C.~P.~Kao, Phys.\ Rev.\  D {\bf 54} (1996) 5636  [hep-ph/9608466];
  T.~M.~Aliev, A.~Ozpineci and M.~Savci, Phys.\ Rev.\  D {\bf 56} (1997) 4260  [hep-ph/9612480];
  D.~Melikhov, N.~Nikitin and S.~Simula, Phys.\ Lett.\  B {\bf 430} (1998) 332  [hep-ph/9803343];
  A.~Ali, G.~Kramer, G~H~Zhu, Eur.\ Phys.\ J.\ C {\bf 47} (2006) 625  [hep-ph/0601034];
  C.~Bobeth, G.~Hiller, G.~Piranishvili, JHEP {\bf 0807} (2008) 106  [arXiv:0805.2525 [hep-ph]].
  G.~Burdman, Phys.\ Rev.\  D {\bf 57} (1998) 4254 [hep-ph/9710550].
  M.~Beneke, T.~Feldmann and D.~Seidel, Nucl.\ Phys.\  B {\bf 612} (2001) 25 [hep-ph/0106067].
  W.~Altmannshofer, P.~Ball, A.~Bharucha, A.~J.~Buras, D.~M.~Straub and M.~Wick, JHEP {\bf 0901} (2009) 019 [arXiv:0811.1214 [hep-ph]].
\bibitem{Ali:1999mm}
  A.~Ali, P.~Ball, L.~T.~Handoko and G.~Hiller, Phys.\ Rev.\  D {\bf 61} (2000) 074024  [hep-ph/9910221];

\bibitem{NP_ANA}
  For instance:
  C.~Greub, A.~Ioannissian and D.~Wyler, Phys.\ Lett.\ B {\bf 346} (1995) 149  [hep-ph/9408382];
  J.~L.~Hewett and J.~D.~Wells, Phys.\ Rev.\  D {\bf 55} (1997) 5549  [hep-ph/9610323];
  A.~Ali, E.~Lunghi, C.~Greub,G.~Hiller, Phys.\ Rev.\  D {\bf 66} (2002) 034002  [hep-ph/0112300];
  C.~H.~Chen and C.~Q.~Geng, Phys.\ Rev.\  D {\bf 66} (2002) 094018  [hep-ph/0209352];
  T.~Feldmann and J.~Matias, JHEP {\bf 0301} (2003) 074  [hep-ph/0212158];
  G.~Hiller, F.~Kruger, Phys.\ Rev.\  D {\bf 69} (2004) 074020  [hep-ph/0310219];
  F.~Kruger and J.~Matias, Phys.\ Rev.\  D {\bf 71} (2005) 094009  [hep-ph/0502060];
  Y.~G.~Xu, R.~M.~Wang and Y.~D.~Yang, Phys.\ Rev.\  D {\bf 74} (2006) 114019  [hep-ph/0610338];
  A.~Hovhannisyan, W.~S.~Hou and N.~Mahajan, Phys.\ Rev.\  D {\bf 77} (2008) 014016
  [hep-ph/0701046];
  Q.~Chang, X.~Q.~Li and Y.~D.~Yang,  JHEP {\bf 1004}, 052 (2010)  [arXiv:1002.2758 [hep-ph]].

\bibitem{ued}
  P.~Colangelo, F.~De Fazio, R.~Ferrandes and T.~N.~Pham, Phys.\ Rev.\  D {\bf 73} (2006) 115006  [hep-ph/0604029];

\bibitem{CDFPhill}
 CDF Collaboration,``Measurement of forward-backward asymmetry in $B\to K^{(\star)}\mu^+\mu^-$ and firtst observation of $B_s^0\to \phi\mu^+\mu^-$'',
 CDF Note 10047, June 1, 2010, arXiv:1001.1028 [hep-ex].

\bibitem{bsphill}
 T. M. Aliev, M. Savci, Phys. Lett. {\bf B481} (2000) 275;
 T. M. Aliev, M. K. Cakmak, A. Ozpineci, M. Savci, Phys. Rev.  {\bf D64} (2001) 055007;
 T. M. Aliev, M. K. Cakmak, M. Savci, Phys. Nucl. Phys. {\bf B607} (2001) 3005;
 S. Rai Choudhury, N. Gaur, N. Mahajan, Phys. Rev {\bf D66} (2002) 054003;
 T. M. Aliev, A. Ozpineci, M. Savci, Phys. Rev {\bf D67} (2003) 035007;
 A. S. Cornell, N. Gaur, JHEP, 0502:005 (2005); U. O.
 Yilmaz, G. Turan, Eur. Phys. J.  {\bf C51}, 63 (2007);
 Yilmaz, G. Turan, Eur. Phys. J.  {\bf C58}, 555 (2008);
 G. Erkol, G. Turan, Eur. Phys. J. {\bf C25} (2002) 575;
 E. Lunghi, A. Soni, JHEP 1011:121,2010;
 Q. Chang, Y.H Gao,   Nucl. Phys. {\bf B845},179-189,(2011).

\bibitem {bsphiued}
 R. Mohanta, A. K. Giri, Phys. Rev {\bf D75} (2007) 035008.

\bibitem {Ball1998}
P. Ball, W. M. Braun, Phys. Rev {\bf D58} (1998) 094016.

\bibitem {Ball2005}
P. Ball, R. Zwicky, Phys. Rev {\bf D71} (2005) 014029.

\bibitem {Wu2006}
Y. L. Wu, M. Zhong, Y.B. Zuo, Int. J. Mod. Phys. {\bf A21}, (2006) 6125.

\bibitem{Wang:2007an}
  W.~Wang, R.~H.~Li and C.~D.~Lu,
  arXiv:0711.0432 [hep-ph].


\bibitem{Melikov2000}
D. Melikhov, B. Stech, Phys. Rev. {\bf D62} (2000) 014006.

\bibitem{Deandrea}
A. Deandrea, A. D. Polosa, Phys. Rev {\bf D64} (2001) 074012.

\bibitem{Geng2003}
C. Q. Geng, C. C. Liu, J. Phys.  {\bf G29} (2003) 1103.

\bibitem{arkani}
N. Arkani, S. Dimopoulos, G. Dvali, Phys. Lett. B 429, 263 (1998);
Phys. Rev. D 59, 086004 (1999); \\
I. Antoniadis, N. Arkani, S. Dimopoulos, G. Dvali, Phys. Lett. B
439, 257 (1998);\\
I. Antoniadis, Phys. Lett. B 246, 377 (1990);

\bibitem{Randall:1999ee}
  L.~Randall and R.~Sundrum,
  Phys.\ Rev.\ Lett.\  {\bf 83}, 3370 (1999)
  [arXiv:hep-ph/9905221].


\bibitem{ACD}
T. Appelquist, H. C. Cheng, B. A. Dobrescu, Phys. Rev. D 64, 035002 (2001).

\bibitem{buras}
A. J. Buras, M.Spranger,  A. Weiler,  Nucl. Phys. B 660, (2003)
225; \\
A. J. Buras, A. Poschenrieder, M. Spranger,  A. Weiler, Nucl.
Phys., B 678,(2004) 455.


\bibitem{Haisch:2007vb}
  U.~Haisch and A.~Weiler,
  Phys.\ Rev.\  D {\bf 76}, 034014 (2007)
  [arXiv:hep-ph/0703064].


\bibitem{azizi1}
V. Bashiry, M. Bayar, K. Azizi,  Phys. Rev. D 78, 035010 (2008).

\bibitem{colangelo}
P.~Colangelo, F.~De Fazio, R.~Ferrandes and T.~N.~Pham,
  Phys.\ Rev.\  D {\bf 77}, 055019 (2008);\\
M.V. Carlucci, P. Colangelo, F. De Fazio, Phys. Rev. D 80, 055023
(2009).

\bibitem{aliev}
T. M. Aliev, M. Savci,  B. B. Sirvanli,  Eur. Phys. J. C 52, 375 (2007).

\bibitem{aslambey}
 I.  Ahmed, M. A. Paracha, M. J. Aslam, Eur. Phys. J. C 54, 591 (2008).

\bibitem{Altmannshofer:2008dz}
  W.~Altmannshofer, P.~Ball, A.~Bharucha, A.~J.~Buras, D.~M.~Straub and M.~Wick, JHEP {\bf 0901} (2009) 019, arXiv:0811.1214 [hep-ph].

\bibitem{Chetyrkin:1996vx}
  K.~G.~Chetyrkin, M.~Misiak and M.~Munz, Phys.\ Lett.\  B {\bf 400} (1997) 206 [Erratum-ibid.\  B {\bf 425} (1998) 414] [hep-ph/9612313].

\bibitem{Beneke:2001at}
  M.~Beneke, T.~Feldmann and D.~Seidel, Nucl.\ Phys.\  B {\bf 612} (2001) 25 [hep-ph/0106067].

\bibitem{bobeth}
  C.~Bobeth, M.~Misiak and J.~Urban, Nucl.\ Phys.\  B {\bf 574} (2000) 291 [hep-ph/9910220].

\bibitem{bobeth02}
  C.~Bobeth, A.~J.~Buras, F.~Kr\"uger and J.~Urban, Nucl.\ Phys.\  B {\bf 630} (2002) 87 [hep-ph/0112305].

\bibitem{Huber:2005ig}
  T.~Huber, E.~Lunghi, M.~Misiak and D.~Wyler, Nucl.\ Phys.\  B {\bf 740} (2006) 105 [hep-ph/0512066].

\bibitem{Buras:1993xp}
  A.~J.~Buras, M.~Misiak, M.~Munz and S.~Pokorski, Nucl.\ Phys.\  B {\bf 424} (1994) 374 [hep-ph/9311345].

\end{thebibliography}
\end{document}